\documentclass[twocolumn,superscriptaddress,prb,showpacs]{revtex4-1}

\usepackage{amsmath,amssymb}
\usepackage{graphicx}
\usepackage{bm}
\usepackage[hypertex]{hyperref}
\usepackage{srcltx} 
\allowdisplaybreaks

\begin{document}

\title{
Disorder effects on thermal transport on the surface of topological
superconductors \\
by the self-consistent Born approximation
}
\date{\today}
\author{Ryota Nakai}
\affiliation{WPI-Advanced Institute for Materials Research (WPI-AIMR), Tohoku University,
Sendai 980-8577, Japan}
\email{rnakai@wpi-aimr.tohoku.ac.jp}

\author{Kentaro Nomura}
\affiliation{Institute for Materials Research, 
Tohoku University, Sendai 980-8577, Japan}

\begin{abstract}
 We have studied the longitudinal thermal conductivity of the surface of
 a three-dimensional time-reversal symmetric topological
 superconductor with random disorder.
Majorana fermions on the surface of  topological superconductors
have a response to the gravitational field,
which is realized as a thermal response to the temperature gradient
inside of the material. 
Because of the presence of both time-reversal symmetry and particle-hole symmetry,
disorder on the surface  emerges in the Hamiltonian 
only as spatial deformations of the pair potential.
In terms of the gravitational field,
the  disorder results in spatial fluctuations of the metric.
We consider  disorder effects on the thermal conductivity perturbatively
within the self-consistent Born approximation.
The density of states is calculated with the Green's function technique
and the thermal conductivity of the surface modes is derived through
 the electronic conductivity using Wiedemann-Franz law for the Majorana fermions.
\end{abstract}


\maketitle

\section{Introduction}

The concept of the topological phase
provides a new class of quantum states of matter where
 the conventional understanding of phases due to the spontaneous symmetry
 breaking can not be applied.
A topological phase and a trivial gapped phase are separated by 
a quantum critical point,
and they can not be continuously connected without closing the energy gap.
As for non-interacting fermions,
the topological insulator is a topological phase of the electron system
and the topological superconductor is that of the system of 
Bogoliubov quasi-particles in the BCS mean field theory of
superconductors\cite{hasan10,qi11}. 
Both topological phases are distinguished from usual insulator and
superconductor phases
by the topology of the occupied Bloch bands\cite{thouless82}.

The topological phases of non-interacting fermionic systems are 
classified by discrete symmetries such as
time-reversal symmetry, particle-hole symmetry, chiral symmetry,
reflection symmetry, point group symmetries, and so on.
The former three symmetries were first brought into 
the classification of topological insulators and
topological superconductors\cite{schnyder08,kitaev09,ryu10} 
according to the Altland-Zirnbauer symmetry classes\cite{altland97}.
The topological phases appear in three spatial dimensions
in five out of the ten of Altland-Zirnbauer symmetry classes.
In symmetry class AII,
the three-dimensional $\mathbb{Z}_2$ topological insulator has been theoretically
predicted\cite{liang07,moore07,roy09}
and experimentally examined\cite{hsieh08}.
In symmetry class DIII,  the $^3$He-B phase is considered to
be a realization of a topological superfluid phase,
and Cu-doped Bi$_2$Se$_3$ has been proposed to be a candidate material
of topological superconductors\cite{hor10,fu10}.

One of the intriguing properties of the topological insulators 
and the topological superconductors is
emergence of gapless surface modes,
while the bulk is fully gapped as in usual insulators and superconductors.
There is one-to-one correspondence between 
nontrivial bulk topological numbers and the
appearance of surface modes (bulk-boundary correspondence),
and symmetries of the bulk topological phases protect the surface
modes from opening a gap and localization.

The surface of the topological superconductor hosts the Majorana fermion
whose anti-particle is itself.
The topological superconductor and superfluid in symmetry class DIII
have surface Majorana fermions in a time-reversal pair.
Robustness of the surface Majorana fermion modes in symmetry class DIII
is partially understood as follows.
Under both time-reversal symmetry and particle-hole symmetry,
only the variations of the Fermi velocity of the Dirac equation are
allowed as perturbations,
and thus the perturbations can not open an energy gap.
These types of perturbations are described in terms of the gravitational
field\cite{volovik08} which couples to the Majorana fermions.
While charge neutrality of the Majorana particles prevents direct
detection by an external electromagnetic field,
transport phenomena of the surface Majorana fermions could be
observed by applying a gravitational field, 
realizations of which are 
temperature gradients and rotations inside of the material\cite{luttinger64}.

The electronic conductivity of the Dirac (Majorana) fermion systems
at the exact zero Fermi energy
is known to be of the order of a universal constant $e^2/h$, 
which is referred to as the minimal conductivity\cite{fradkin86,lee93,ludwig94,ryu07,ziegler07}.
The surface of topological insulators and topological
superconductors naturally realize this situation.
The relation between the electronic conductivity and 
the thermal conductivity is given by the Wiedemann-Franz law\cite{smrcka77}. 
The Wiedemann-Franz law for Majorana particles\cite{nomura12} reads as
\begin{align}
 \kappa=\frac{\pi^2k_B^2T}{6e^2}\sigma.
\end{align}
Note that the electronic conductivity of charge neutral Majorana fermions means
just the response function derived from the Kubo formula,
the calculation of which is not restricted to charged particles.
The coefficient of the right-hand side of the above law
 is half of that of the complex fermion case, since a Majorana (real) fermion 
is equal to a half of a complex fermion.

In this paper,
static and dynamical properties of the surface Majorana fermions of the topological
superconductor in symmetry class DIII with disorder are studied through the Green's function
technique\cite{shon98}.
With respect to small deformations of the pair potential,
a part of the perturbation series
that is significant in the conductive regime
is picked up by the self-consistent Born approximation (SCBA).
This paper is organized as follows.
In section 2,
the model Hamiltonian of the three-dimensional 
topological superconductor in  symmetry class DIII 
is introduced with its defining symmetries.
Possible terms that are allowed to be added to the non-perturbed
Hamiltonian under two discrete symmetries 
and the perturbation terms that we consider in this paper are shown.
In section 3,
the disorder averaged Green's function is derived by solving equations
of the SCBA.
The density of states is obtained through the averaged Green's function.
In section 4,
the electronic conductivity is derived by summing up infinite series of
perturbation terms of the SCBA,
and the thermal conductivity is obtained via the Wiedemann-Franz law.

\section{Surface states of topological superconductors in class DIII}

In this section,
the low-energy effective Hamiltonian of 
the three-dimensional topological superconductor\cite{schnyder08,kitaev09,ryu10} 
in symmetry class DIII\cite{altland97} is reviewed 
with its symmetry classification.
The surface of topological superconductors hosts
two-dimensional Majorana surface modes.
We will show that
characterizing symmetries of the symmetry class DIII 
puts constraints on possible perturbations on the surface modes.
Disorder on the surface is then regarded
as spatial deformations of the pair potential.

\subsection{Model Hamiltonian of topological superconductors in class DIII}

Symmetry classes of the random matrix theory are brought to 
classification of 
the topological insulators and the topological superconductors,
which are defined with three types of discrete symmetries: 
time-reversal symmetry, particle-hole symmetry,
and a combination of them called chiral symmetry.
Symmetry class DIII corresponds to time-reversal-symmetric
superconductors with the triplet Cooper pairing.
We consider a  bulk $4\times 4$ matrix Hamiltonian 
in the momentum space 
$H_{\text{bulk}}(\bm{k})$.
Four fermion flavors in the Hamiltonian are the product of
two flavors of spin up and down, and two flavors of the particle and
the hole. 
Time-reversal symmetry for spin 1/2 is defined by
\begin{align}
 (is^y)H^{T}_{\text{bulk}}(-\bm{k})(-is^y)
 =
 H_{\text{bulk}}(\bm{k}), \label{eq:time-reversal}
\end{align}
where $s^i (i=x,y,z)$ is a set of the Pauli matrix 
for the spin degrees of freedom and
$T$ denotes the transpose of matrices.
The square of the time-reversal operation is
$(is^y K)^2=-1$, where $K$ denotes the complex conjugation operator.
Particle-hole symmetry for triplet superconductors is defined by
\begin{align}
 t^xH_{\text{bulk}}^T(-\bm{k})t^x
=
-H_{\text{bulk}}(\bm{k}), \label{eq:particle-hole}
\end{align}
where $t^i (i=x,y,z)$ is a set of the Pauli matrix
for the particle-hole degrees of freedom.
The square of the particle-hole conjugation operator is
$(t^x K)^2=+1$.

As a low-energy effective model of topological superconductors,
we consider the Bogoliubov-de Gennes Hamiltonian
of the form
\begin{align}
 \mathcal{H}_{\text{bulk}}
 =
 \frac{1}{2}
 \sum_{k}
 \Psi_{k}^{\dagger}
 H_{\text{bulk}}(\bm{k})
 \Psi_{k},
\end{align}
where
$\Psi_k=(c_{k\uparrow},
c_{k\downarrow},
c^{\dagger}_{-k\uparrow},
c^{\dagger}_{-k\downarrow})^T$
($c_{ks}$ and $c^{\dagger}_{ks}$ are the creation and the annihilation
operators of a complex fermion with the momentum $k$ and the spin $s$).
From Hermiticity and particle-hole symmetry,
the matrix elements of the Hamiltonian are written as
\begin{align}
 H_{\text{bulk}}(\bm{k}) 
 =
 \begin{pmatrix}
  \Xi_k & \Delta_k \\
  \Delta_k^{\dagger} & -\Xi_{-k}^T
 \end{pmatrix}, \label{eq:Hamiltonian_matrix_DIII}
\end{align}
where $\Xi_k$ and $\Delta_k$ are $2\times 2$ matrices satisfying
$\Delta_{-k}^T=-\Delta_k$.
Time reversal symmetry for $\Xi_k$ and $\Delta_k$ is
given by $is^y\Xi_{-k}^T(-is^y)=\Xi_k$ and $is^y\Delta_{-k}^T(-is^y)=\Delta_k$.
We consider the forms of $\Xi_k$ and $\Delta_k$ given by
\begin{align}
 &\Xi_k=\Xi_{-k}^T
 =
 \left(
 \frac{\hbar^2k^2}{2m}-\mu
 \right)s^0, \label{eq:diagonal_kinetic}\\
 &\Delta_k
 =
 \Delta_0 \bm{k}\cdot \bm{s} (is^y)
 =
 \Delta_0
 \begin{pmatrix}
  -k_x+ik_y & k_z \\
  k_z & k_x+ik_y
 \end{pmatrix}, \label{eq:off-diagonal_gap}
\end{align}
which satisfy all the symmetries required for the Hamiltonian in
symmetry class DIII.
Note that we only consider
real and positive $\Delta_0$ without loss of generality since the
phase factor of $\Delta_0$ can be removed by 
a U(1) gauge transformation of the fermion operators.
The Hamiltonian (\ref{eq:Hamiltonian_matrix_DIII}) represents, for example,
the B phase of superfluid $^3$He\cite{volovik03}.

\subsection{Surface states of topological superconductor}

Since 
the kinetic energy in the diagonal blocks of the Hamiltonian matrix (\ref{eq:diagonal_kinetic})
can be ignored in the long wave-length limit,
the diagonal blocks are regarded as a mass term 
with respect to the Dirac cone structure of the off-diagonal 
pair potential term (\ref{eq:off-diagonal_gap}).
The boundary of a material is, for convenience, 
defined by a position dependent mass term $\mu(z)$,
where the $z$ direction is normal to the surface of the material.
Putting a material in the region of $z<0$,
$\mu$ smoothly changes its sign from negative to positive
when moving from the material to the vacuum (outside of the material),
and converges to a finite value away from the surface:
\begin{align}
 \mu(z)
 \to
 \left\{
 \begin{array}{ll}
  \mu\quad &(z\to\infty)\\
  -\mu\quad &(z\to -\infty)
 \end{array}
 \right. ,
\end{align}
where $\mu>0$.

In the coordinate space description, the bulk Hamiltonian is written as
\begin{align}
 \mathcal{H}_{\text{bulk}}
 =
 \frac{1}{2}
 \int d^3r
 \Psi^{\dagger}(\bm{r}) H_{\text{bulk}}(\bm{r}) \Psi(\bm{r}),
\end{align}
where 
\begin{align}
 H_{\text{bulk}}(\bm{r})
 &=
 \begin{pmatrix}
  -\mu(z)s^0 & \Delta_0 (-i\bm{\partial})\cdot \bm{s}(is^y) \\
  \Delta_0 (-is^y)(-i\bm{\partial})\cdot \bm{s} & \mu(z)s^0
 \end{pmatrix} \notag\\
 &=
  -i\Delta_0(
  -\partial_x s^z\otimes t^x 
  -\partial_y s^0\otimes t^y 
  +\partial_z s^x\otimes t^x) \notag\\
  &\quad\,
  -\mu(z) s^0\otimes t^z.
  \label{eq:Hamiltonian_near_boundary}
\end{align}
Using the fermion operator in the coordinate space 
$c_s(\bm{r})=(1/L)\sum_k e^{i\bm{k}\cdot \bm{r}}c_{ks}$,
the spinor is $\Psi(\bm{r})=(c_{\uparrow}(\bm{r}),
c_{\downarrow}(\bm{r}),
c^{\dagger}_{\uparrow}(\bm{r}),
c^{\dagger}_{\downarrow}(\bm{r}))$.
The eigenfunctions of the Hamiltonian
(\ref{eq:Hamiltonian_near_boundary}) 
near the surface
are the product of the plane waves in surface direction ($x$- and $y$-direction),
and a bound function normal to the surface ($z$-direction): 
$e^{i(k_xx+k_yy)}\psi^{1(2)}(z)$,
with
\begin{align}
 \psi^{1(2)}(z)=
 \exp
 \left[
 -\int^z dz' \frac{\mu(z')}{\Delta_0}
 \right]
 |1(2)\rangle. \label{eq:surface_eigenfunction}
\end{align}
Four-component spinors $|1\rangle$ and $|2\rangle$
are the basis vectors that span the eigenspace
of $(s^x\otimes t^y)$ with the eigenvalue $-1$
which are assigned to bound functions,
while those with the eigenvalue $+1$  are diverging functions 
which cannot be normalized.

The Hamiltonian for surface modes 
is reduced from the bulk Hamiltonian
(\ref{eq:Hamiltonian_near_boundary}) by 
projecting the Hilbert space 
onto the subspace spanned by
the product of $\psi^{1(2)}(z)$ and functions of $x$ and $y$.
Through this process,
the Hamiltonian becomes independent of $z$,
and the four-component spinor degrees of freedom
are reduced to the two-component ones.
We use the basis vectors $|1(2)\rangle$ as
\begin{align}
 |1\rangle=
 \begin{pmatrix}
  i \\ 1 \\ -i \\ 1
 \end{pmatrix}
 /2,
 \quad
 |2\rangle=
 \begin{pmatrix}
  1 \\ i \\ 1 \\ -i
 \end{pmatrix}
 /2.
\end{align}
With these vectors,
operators of the surface Majorana fermions at the  position $(x,y)$ are  given by 
$\gamma^{1(2)}(x,y)\propto\int dz{\psi^{1(2)}}^{\dagger}(z)\Psi(\bm{r})$,
or explicitly,
\begin{align}
 \gamma^{1}
 &\propto
 \int dz \exp
 \left[
 -\int^z dz' \frac{\mu(z')}{\Delta_0}
 \right]
 (\gamma_{\uparrow 2}+\gamma_{\downarrow 1}),
 \\
 \gamma^{2}
 &\propto
 \int dz \exp
 \left[
 -\int^z dz' \frac{\mu(z')}{\Delta_0}
 \right]
 (\gamma_{\uparrow 1}+\gamma_{\downarrow 2}),
\end{align}
where $\gamma_{s1}=c_{s}(\bm{r})+c_s^{\dagger}(\bm{r})$
and $\gamma_{s2}=(c_{s}(\bm{r})-c_s^{\dagger}(\bm{r}))/i$ 
are Majorana fermion operators generated from
complex fermion operators $c_{s}$ with spin $s=\uparrow, \downarrow$.
Obviously,
the operators $\gamma^1$ and $\gamma^2$ satisfy
the Majorana condition $\gamma^i={\gamma^i}^{\dagger}$.
The wavefunctions of the surface mode can be written as 
$u(x,y)\psi^1(z)+v(x,y)\psi^2(z)$.
Therefore the Hamiltonian for the surface modes is given by
\begin{align}
 H_0(\bm{r})=-i\Delta_0(\partial_x\sigma^z+\partial_y\sigma^x), \label{eq:Hamiltonian_surface}
\end{align}
where $\sigma^i (i=x,y)$ is the set of the Pauli matrix 
for the two-component spinor $(u(x,y),v(x,y))^T$.
After projecting onto the surface modes,
the time-reversal operator becomes $T=i\sigma^y K$,
and the particle-hole conjugation operator becomes $C=K$.
In the following,
we use the notation $(\tilde{\sigma}^x,\tilde{\sigma}^y)$ in place of the Pauli matrix 
$(\sigma^z,\sigma^x)$ for convenience,
and the Hamiltonian (\ref{eq:Hamiltonian_surface}) is briefly rewritten as 
$H_0(\bm{r})=-i\Delta_0\bm{\partial}\cdot\tilde{\bm{\sigma}}$.
Implicitly $\sigma^y$ is replaced by $\tilde{\sigma^z}$ accordingly,
while $\sigma^0$ is unchanged.
This replacement does not affect the following calculations since the
algebra that the Pauli matrix obeys is invariant under this
replacement. 
Also the physical meaning of perturbation terms added to the surface
Hamiltonian (\ref{eq:Hamiltonian_surface}), like the chemical
potential term and the mass term, is conserved under the replacement 
since their meaning is dependent on the explicit form of
the unperturbed Dirac Hamiltonian.

\subsection{Deformation of the pair potential}

Under both of  time-reversal symmetry and  
particle-hole symmetry,
the possible perturbation terms that are allowed to be added to 
the surface Hamiltonian (\ref{eq:Hamiltonian_surface})
are strictly limited.
Prohibited terms are, for example,
a chemical potential $\mu \sigma^0$, which breaks particle-hole symmetry,
a mass term $m\tilde{\sigma}^z$, 
which breaks time-reversal symmetry,
and U(1) gauge potential terms $A_x\tilde{\sigma}^x, A_y\tilde{\sigma}^y$, 
which break both of them.
Note that symmetries of each perturbation term must be examined
before replacing the Pauli matrix since the action of the
time-reversal and particle-hole conjugation operators is dependent
on the elements of the $2\times 2$ Pauli matrix.
In the momentum-space representation, 
the available terms are ones listed in the following:
\begin{align}
 &(\text{odd function of }k_x, k_y)\times\tilde{\sigma}^x, \notag\\
 &(\text{odd function of }k_x, k_y)\times\tilde{\sigma}^y.
\end{align}
Among them, only the terms proportional to $k_x$ or $k_y$ have significant contributions,
since higher order terms can be neglected in the long wave-length limit.
The full Hamiltonian that we consider in this paper is as follows:
\begin{align}
 H=H_0+U=\frac{-i}{2}\{\Delta(\bm{r}),\partial_x\tilde{\sigma}^x+\partial_y\tilde{\sigma}^y\}. 
 \label{eq:Hamiltonian_surface_whole}
\end{align}
The Hamiltonian (\ref{eq:Hamiltonian_surface_whole}) means that
the pair potential is spatially deformed by random disorders,
which will be explained in the following.
A deformation of the pair potential
represented by a small conformal factor $\Lambda(\bm{r})(\ll 1)$ as
$\Delta(\bm{r})=\Delta_0e^{\Lambda(\bm{r})}$
can be undertaken by the Pauli matrix with the vierbein field 
$e^a_i(\bm{r})=\delta^a_{i}e^{\Lambda(\bm{r})}$ 
as $\tilde{\sigma}^a(\bm{r})=e^a_i(\bm{r})\tilde{\sigma}^i$.
Thus the Hamiltonian (\ref{eq:Hamiltonian_surface_whole}) 
describes the surface Majorana fermions on the curved space with the metric 
\begin{align}
 g^{ab}(\bm{r})&=\{\tilde{\sigma}^a(\bm{r}),\tilde{\sigma}^b(\bm{r})\} \notag\\
 &=e^{2\Lambda(\bm{r})}\delta^{ab}, \label{eq:metric}
\end{align}
where $a,b$ are indices of spatial coordinates.
The metric (\ref{eq:metric}) indicate that
the Hamiltonian (\ref{eq:Hamiltonian_surface_whole}) is continuously 
connected from the unperturbed Hamiltonian
(\ref{eq:Hamiltonian_surface}) 
by a conformal transformation of the spatial coordinates.
The perturbation term is given by
\begin{align}
 U(\bm{r})
 \simeq
 \frac{-i\Delta_0}{2}\{\Lambda(\bm{r}),\partial_x\tilde{\sigma}^x+\partial_y\tilde{\sigma}^y\},
 \label{eq:Hamiltonian_surface_perturbation}
\end{align}
where all terms second order in $\Lambda(\bm{r})$ or higher are neglected. 
We should note that the spin connection does not contribute to the
 Hamiltonian for the case of a single Dirac cone Hamiltonian 
(Hamiltonian described by the $2\times 2$ Pauli matrices)\cite{nakahara90}.

Consider situation where
time-reversal symmetry preserved scatterers are contained
 inside  the topological superconductor.
Only scatterers near the surface have an influence
on the surface modes.
Point-like scatterers are randomly distributed on the surface,
and a single scatterer 
at the position $\bm{r}_i=(x_i,y_i)\,(i=1,2,\cdots,N_{\text{i}})$
affects the pair potential around $\bm{r}_i$
as the form of Gaussian
\begin{align}
 \Lambda(\bm{r})=-c\sum_{i=1}^{N_{\text{i}}} e^{-|\bm{r}-\bm{r}_i|^2/2R^2},
 \label{eq:conformal_factor}
\end{align}
where $c$ is a dimensionless parameter that represents the 
strength of disorder,
$R$ is an effective range of the influence of a scatterer,
both of which are assumed to be common for all scatterers,
and $N_{\text{i}}$ is the number of the of scatterers.
We consider the effective range $R$ to be much larger than the lattice
spacing so as to justify the continuous description of 
the conformal factor (\ref{eq:conformal_factor}).
By the Fourier transformation,
the conformal factor in the momentum space is 
\begin{align}
 \Lambda(\bm{k})
 &=
 \frac{1}{L}\int d^2r e^{i\bm{k}\cdot \bm{r}}\Lambda(\bm{r}) \notag\\
 &=
 -\frac{2\pi cR^2}{L}\sum_{i=1}^{N_{\text{i}}}
 e^{-R^2|\bm{k}|^2/2+i\bm{k}\cdot \bm{r}_i}.
\end{align}

On the surface of the topological superconductor,
distribution of the positions of the scatterers is random in two-dimensional space.
The average of physical quantities over the position of the scatterers
$\bm{r}_i$ is denoted by $\langle A\rangle$, 
and its definition is 
\begin{align}
 \langle A\rangle=\left[\prod_{i=1}^{N_{\text{i}}}\frac{1}{L^2}\int d^2r_i \right]A.
\end{align}
First, the disorder average of a single conformal
factor gives $\langle\Lambda(\bm{k})\rangle=2\pi N_{\text{i}}cR^2\delta_{k,0}/L$.
Next  the disorder average of the product of two conformal factors,
that is, the correlator of conformal factors is 
\begin{align}
 \langle \Lambda(\bm{k})\Lambda(\bm{k}')\rangle 
 &=
 \delta_{k+k',0}n_{\text{i}}(2\pi cR^2)^2e^{-R^2|\bm{k}|^2},
\end{align}
where $n_{\text{i}}=N_{\text{i}}/L^2$ is the number of scatterers per
unit area.
For further simplification,
we assume the effective range of a single scatterer $R$ to be infinity.
In this long-ranged limit,
the correlator of the conformal factor becomes 
\begin{align}
 \langle \Lambda(\bm{k})\Lambda(\bm{k}')\rangle 
 &\to
 \delta_{k+k',0}n_{\text{i}}(2\pi cR^2)^2 \delta_{k,0} \notag\\
 &\equiv \delta_{k+k',0}L^2v_{\text{i}}\delta_{k,0}.
 \label{eq:conformal-factor_correlation}
\end{align}
where $v_{\text{i}}= n_{\text{i}}(2\pi cR^2)^2/L^2$
 is a parameter that represents the intensity of disorder.

\section{Averaged Green's function by the self-consistent Born approximation\label{sec:scB_approx}}

In this section,
the disorder effects on the surface Majorana fermions
are studied through the disorder-averaged Green's function
\begin{align}
 \tilde{G}(\mu)
 \equiv
 \langle G(\mu)\rangle
 =
 \left\langle
 \frac{1}{\mu-H}
 \right\rangle
 \label{eq:averaged_GF}
\end{align}
within the SCBA\cite{shon98}.
Note that, in (\ref{eq:averaged_GF}) and hereafter,
$\sigma^0$ is not written explicitly.
For the long-ranged limit of the deformations of the
pair potential,
resultant difference from the Green's function in the clean limit is 
fully described by a single parameter $A$.
The density of state is obtained by the averaged Green's function.

\subsection{Self-consistent Born approximation}

Consider the situation that the deformations of the pair potential is
much smaller than the original superconducting energy gap ($\Lambda(\bm{r})\ll 1$).
The disorder term (\ref{eq:Hamiltonian_surface_perturbation}) is
treated perturbatively
with respect to the bare Green's function
\begin{align}
 G_0(\mu)=\frac{1}{\mu-H_0}.
\end{align}
The self-energy $\Sigma(\mu)$ is introduced 
by the Dyson's equation
\begin{align}
 \tilde{G}(\mu)=G_0(\mu)+G_0(\mu)\Sigma(\mu)\tilde{G}(\mu).
 \label{eq:Dyson_equation}
\end{align}
Here, we should note that although 
the nonzero Fermi energy term is prohibited from the symmetry argument 
in the previous section,
we will relax this condition for a while in order to examine properties away
from $\mu=0$. 
The disorder average of a single $U$ term is omitted
since it only shifts the energy due to the fact
$\langle\Lambda(\bm{k})\rangle=2\pi N_{\text{i}}cR^2\delta_{k,0}/L$,
and also the disorder average of the product of more than three of 
the $U$ term can be neglected when the scatterers are not densely distributed.
The SCBA for the self-energy is given by
\begin{align}
 \Sigma(\mu)=\langle U\tilde{G}(\mu)U\rangle. 
 \label{eq:scBorn_self-energy}
\end{align}
The averaged Green's function and the self-energy
are derived self-consistently
by combining (\ref{eq:scBorn_self-energy}) with the alternative 
representation 
of the definition of the self-energy (\ref{eq:Dyson_equation}) as
\begin{align}
 \tilde{G}(\mu)=\frac{1}{\mu-H_0-\Sigma(\mu)}.
 \label{eq:averaged_GF_self-energy}
\end{align}

We will use the momentum-space representation of the Hamiltonian 
to apply the above method to the current situation, i.e.,
\begin{align}
 \mathcal{H}=\sum_{k,k'} \Psi_k^{\dagger}(H_{0kk'}+U_{kk'})\Psi_{k'},
\end{align}
where the $(\bm{k},\bm{k}')$ component of each term is
\begin{align}
 &H_{0kk'}=\delta_{k,k'}\,v_F \hbar \,\tilde{\bm{\sigma}}\cdot \bm{k}, \\
 &U_{kk'}=\frac{v_F\hbar}{2L}\Lambda(\bm{k}-\bm{k}') \tilde{\bm{\sigma}}\cdot (\bm{k}+\bm{k}'). 
 \label{eq:matrix_element_perturbation}
\end{align}
In the above equation, 
the pair potential $\Delta_0/2$ is replaced by
the Fermi velocity $v_F$ in order to fit
the notation of the Dirac equation.
Note that (\ref{eq:matrix_element_perturbation}) 
satisfies Hermiticity ($U_{kk'}^{\dagger}=U_{k'k}$)
since $\Lambda^{\ast}(-\bm{k})=\Lambda(\bm{k})$.

After averaging over the positions of the scatterers,
the translational invariance is recovered,
and thus the averaged Green's function $\tilde{G}$ and the 
self-energy $\Sigma$ are diagonal with respect to the momentum.
Introducing the averaged Fermi energy $F_0$ and 
the averaged momenta $\bm{F}=(F_x,F_y)$ by 
$\tilde{G}^{-1}=F_0-\tilde{\bm{\sigma}}\cdot \bm{F}$,
the self-energy is given by 
\begin{align}
 \Sigma(\bm{k},\mu)
 &=
 G_{0k}^{-1}(\mu)-\tilde{G}_{k}^{-1}(\mu) \notag\\
 &=
 \mu-F_{0k}-\tilde{\bm{\sigma}}\cdot(v_F\hbar\bm{k}-\bm{F}_k),
 \label{eq:averaged_GF_self-energy2}
\end{align}
where $F_{0k}$ and $\bm{F}_k$ are, respectively,
the $\bm{k}$ component of $F_0$ and that of $\bm{F}$.
Similarly,
by substituting (\ref{eq:conformal-factor_correlation}),
(\ref{eq:scBorn_self-energy}) becomes
\begin{align}
 \Sigma(\bm{k},\mu)
 &=
 \sum_{k'}\langle
 U_{kk'}\tilde{G}_{k'}(\mu)U_{k'k}\rangle \notag\\
 &=
 \frac{v_{\text{i}}(v_F\hbar \tilde{\bm{\sigma}}\cdot \bm{k})(F_{0k}+\tilde{\bm{\sigma}}\cdot
 \bm{F}_{k})(v_F\hbar \tilde{\bm{\sigma}}\cdot \bm{k})} {F_{0k}^2-|\bm{F}_k|^2}.
 \label{eq:scBorn_self-energy2}
\end{align}
Removing the self-energy by equating 
the right-hand side of (\ref{eq:averaged_GF_self-energy2})
and that of (\ref{eq:scBorn_self-energy2}),
and decomposing them into equations proportional to $\sigma^0,\tilde{\sigma}^x$ and
$\tilde{\sigma}^y$,
we obtain a set of self-consistent equations as
\begin{align}
 \left(
 \begin{array}{l}
  \displaystyle
  F_{0k}=\frac{\mu}{1+A} \\
  \displaystyle
  \bm{F}_k=\frac{v_F\hbar\bm{k}}{1-A}
 \end{array}
 \right. ,
 \label{eq:scBorn_A}
\end{align}
where 
$A=v_{\text{i}}(v_F\hbar|\bm{k}|)^2/(F_{0k}^2-|\bm{F}_k|^2)$
is a single parameter that represents the disorder effects calculated within
the SCBA.

\subsection{Solutions of $A$}

The relation between 
the bare and the averaged Green's function (\ref{eq:scBorn_A})
indicates that a couple of self-consistent integral equations of 
the averaged Green's function and 
the self-energy are reduced to algebraic equations of $A$.
The equation of $A$ is given by the definition of $A$ as
\begin{align}
 A
 =
 v_{\text{i}}
 \frac{(v_F\hbar|\bm{k}|)^2}
 {\mu^2/(1+A)^2-(v_F\hbar|\bm{k}|)^2/(1-A)^2}.
 \label{eq:scEquation_A}
\end{align}
Note that 
the equation of $A$ is determined by two parameters,
the intensity of disorder $v_{\text{i}}$
and the ratio of the momentum to the Fermi energy $\kappa=v_F\hbar|\bm{k}|/|\mu|$,
except for a point $\mu=0$ (or equivalently we can consider a
parameter $|\mu|/v_F\hbar|\bm{k}|$ except for $|\bm{k}|=0$).
Then the problem is decomposed into two parts,
one for $\mu\neq 0$ and the other for $\mu=0$.

\begin{figure}
  \includegraphics[width=65mm]{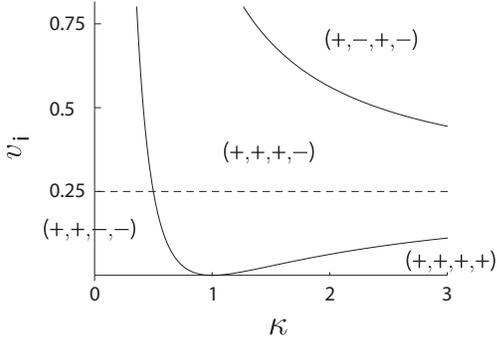}
 \caption{Signs of $l_1, l_2, l_3$, and $l_4$ as functions of 
 $\kappa(=v_F\hbar|\bm{k}|/|\epsilon|)$ and $v_{\text{i}}$
 are shown in $\kappa$-$v_{\text{i}}$ space.
 The signs in the figure indicate
 $(\text{sgn}[l_1],\text{sgn}[l_2],\text{sgn}[l_3],\text{sgn}[l_4])$.
The solid lines partitioning $\kappa$-$v_{\text{i}}$ space 
 are $l_2=0$, $l_3=0$, and $l_4=0$. 
 The dotted line at $v_{\text{i}}=0.25$ is an asymptote of
 $l_2=0$ or $l_3=0$ in the large $\kappa$ region.
 \label{fig:A_region}}
\end{figure}

First, we consider the case $\mu\neq 0$.
The equation (\ref{eq:scEquation_A}) has four branches of solutions.
The explicit forms of the solutions are given by
\begin{align}
 A_{\pm\pm}
 =-
 \left(\frac{l_1^{1/2}-l_2^{1/2}}{l_1^{1/2}+l_2^{1/2}}\right)^{\pm1}
 \left(\frac{l_3^{1/2}-l_4^{1/2}}{l_3^{1/2}+l_4^{1/2}}\right)^{\pm1},
 \label{eq:A_four_branch}
\end{align}
where the two signs in the subscript of $A$ in the left-hand side 
correspond, respectively, to the two signs in right-hand side,
and they can be taken independently.
In the following, the square root of a negative value indicates
a square root that has a positive imaginary part.
The four variables are 
\begin{align}
 \left\{
 \begin{array}{l}
  l_1=(1+2v_{\text{i}}^{1/2})\kappa+1 \\
  l_2=(1-2v_{\text{i}}^{1/2})\kappa+1 \\
  l_3=(1+2v_{\text{i}}^{1/2})\kappa-1 \\
  l_4=(1-2v_{\text{i}}^{1/2})\kappa-1
 \end{array}
 \right. .
 \label{eq:l1_to_l4}
\end{align}
As can be readily seen from the expression (\ref{eq:A_four_branch}),
$A$ is real when the signs of both $l_1\times l_2$ and $l_3\times l_4$ are positive.
Otherwise, that is, when at least either one of the signs of $l_1\times l_2$ and
$l_3\times l_4$ is negative,
$A$ can be complex valued.
The signs of the above four variables are shown in the
$\kappa$-$v_{\text{i}}$ space in Fig.\ref{fig:A_region}.
The boundaries where the sign of one of the variables changes
are $l_2=0$, $l_3=0$ and $l_4=0$.
Thus, the lower region (below $l_4=0$) and 
upper-left region (left side of $l_3=0$) in Fig.\ref{fig:A_region}
have real solutions of $A$,
and in the other regions,
complex solutions of $A$ are realized.
There is an upper bound of the intensity of disorder at
$v_{\text{i}}=0.25$ (shown in Fig.\ref{fig:A_region} by a dotted line)
up to which the solutions of  $A$ are continuously connected from the
clean limit. 
Throughout this paper,
we consider only $v_{\text{i}}<0.25$.

Among the four branches of the solutions of $A$,
only ones that converge to zero in the limit of $v_{\text{i}}\to 0$ make
physical sense,
since $A$ represents deviation from the clean limit.
In the complex $A$ region, however,
we do not impose this condition,
since the complex $A$ region in $\kappa$ space shrinks to a point
$\kappa=1$ in the clean limit,
and the Green's functions have singular behavior there.
When $l_1\times l_2$ and $l_3\times l_4$ are positive, we observe
\begin{align}
 &l_1^{1/2}+l_2^{1/2}\simeq 2(\kappa+1)^{1/2}\sim O(1),\\
 &l_1^{1/2}-l_2^{1/2}\simeq
 \frac{2\kappa}{(\kappa+1)^{1/2}}v_{\text{i}}^{1/2}
 \sim  O(v_{\text{i}}^{1/2}),
\end{align}
and also
\begin{align}
 &l_3^{1/2}+l_4^{1/2}\simeq 2(\kappa-1)^{1/2}\sim O(1),\\
 &l_3^{1/2}-l_4^{1/2}\simeq
 \frac{2\kappa}{(\kappa-1)^{1/2}}v_{\text{i}}^{1/2}
 \sim  O(v_{\text{i}}^{1/2}),
\end{align}
Note again that we took a positive imaginary part branch for the square root of
a negative value.
The appropriate choice is therefore $A_{++}$ for the real $A$ regions.

In the complex $A$ region,
the solutions will be those that are continuously connected to $A_{++}$
at the two boundaries intervening the real and complex $A$ regions,
that is, $l_3=0$ and $l_4=0$.
From (\ref{eq:l1_to_l4}),
an identity $A_{++}=A_{+-}$ holds when $l_3=0$ or $l_4=0$.
Thus, two possibilities arise for the solutions in the complex $A$ region:
$A_{++}$ and $A_{+-}$, which are related by the complex conjugation.
In the following,
we will show that these two solutions correspond to 
the retarded or the advanced averaged Green's functions.

\begin{figure}
 \begin{tabular}{cc}
  \begin{minipage}[c]{42mm}
   \includegraphics[width=42mm]{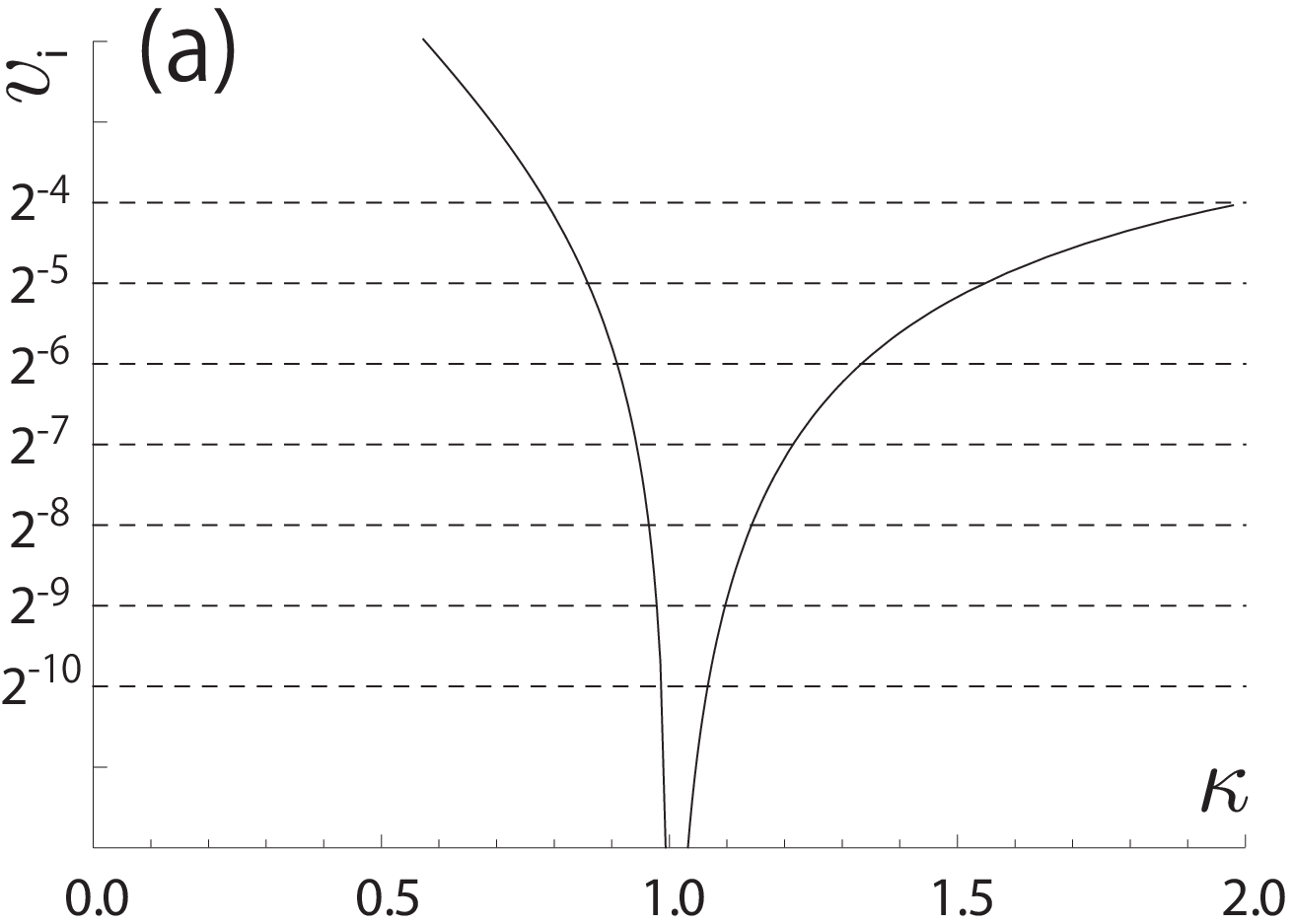}
   \\
   \vspace{4mm}
   \includegraphics[width=42mm]{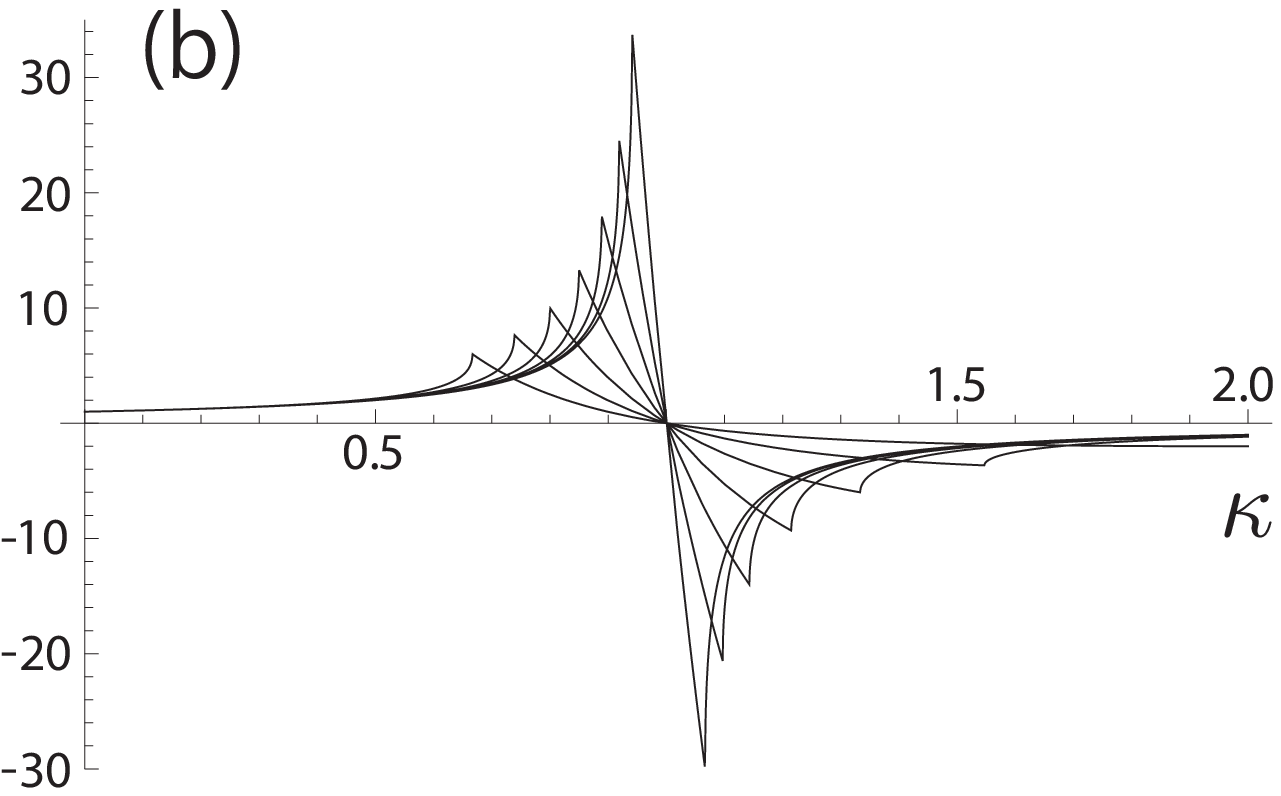}
  \end{minipage}
  & 
  \hspace{2mm}
  \begin{minipage}[c]{40mm}
   \includegraphics[width=40mm]{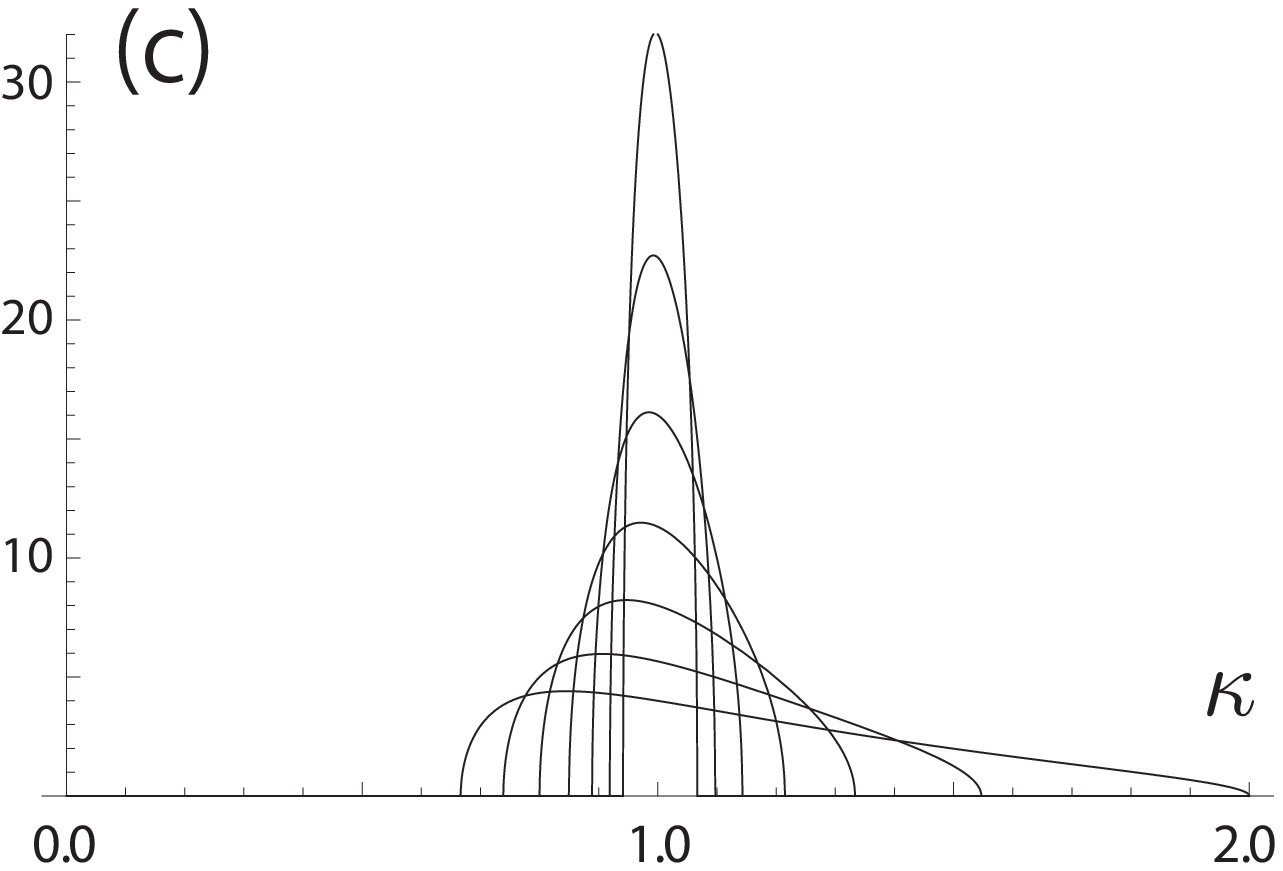}
   \\
   \vspace{3mm}
   \includegraphics[width=40mm]{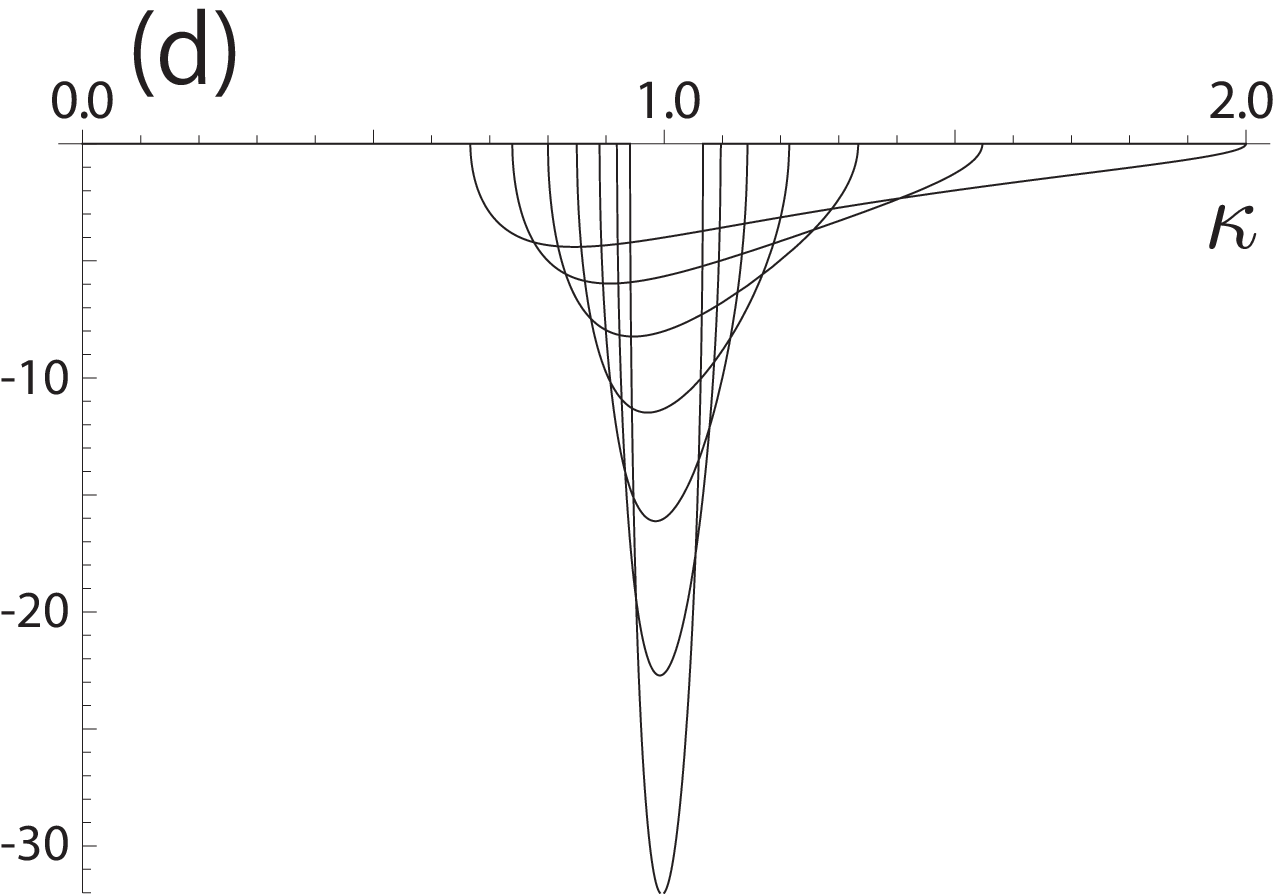}
  \end{minipage}
 \end{tabular}
 \caption{
 One of the eigenvalues of the averaged Green's function
 multiplied by $\mu(>0)$,
 with $v_{\text{i}}=2^{-n}\,(n=4,5,\cdots,10)$ are drawn. 
 (a) The solid lines are boundaries of the solutions of $A$ in $\kappa$-$v_{\text{i}}$
 space, and the dotted lines show the value of $v_{\text{i}}$ which we adopt.
 The other figures correspond to (b) the real part of the averaged Green's function, and the imaginary
 part of (c) the advanced and that of (d) the retarded averaged Green's
 functions as a function of $\kappa=v_F\hbar|\bm{k}|/|\epsilon|$.
 \label{fig:GF}}
\end{figure}

The averaged Green's function is obtained with the solution of $A$ as
\begin{align}
 \tilde{G}_{k}^{-1}(\mu)
 =
 \frac{\mu}{1+A}-\frac{v_F\hbar\tilde{\bm{\sigma}}\cdot \bm{k}}{1-A}.
 \label{eq:averaged_GF_A}
\end{align}
The two branches of the solutions of $A$ in the complex $A$ region,
$A_{++}$ and $A_{+-}$, are
assigned to the retarded or the advanced Green's functions by comparing the
signs of the imaginary part of the inverse of Green's function
between the one in the clean limit and the averaged one.
Here, we consider one of the eigenvalues of the Green's function in which
the sign of the eigenvalue of $\tilde{\bm{\sigma}}\cdot \bm{k}$ is equal to 
the sign of $\mu$ in place of the matrix valued Green's function. 
This side of the eigenvalue has a nonzero imaginary part
when the chemical potential is slightly shifted to the imaginary
direction $\mu\to \mu\pm i\delta$.
The inverse of the bare Green's function has the sign of the imaginary part as follows:
\begin{align}
 &\text{sgn}[\text{Im}[G_{0k}^{-1}(\mu\pm i\delta)]]   
 =\pm 1,
\end{align}
where $\delta$ is a positive infinitesimal parameter.
The sign of the imaginary part of the averaged Green's function
with the complex $A_{+\pm}$ is
\begin{align}
 &\text{sgn}[\text{Im}[\tilde{G}_{k}^{-1}(\mu)]] \notag\\
 &=
 \text{sgn}\left[\text{Im}\left[
 \mu
 \left(
 \frac{1}{1+A_{+\pm}}-\frac{\kappa}{1-A_{+\pm}}
 \right)
 \right]\right] \notag\\
 &=
 \text{sgn}\left[\text{Im}\left[
 -\mu
 \frac{\kappa+1}{2}(l_3^{1/2}\pm l_4^{1/2})^2
 \right]\right] \notag\\
 &=
 \mp \text{sgn}[\mu].
\end{align}
Therefore, the appropriate choices of the branches of $A$ for the 
retarded (denoted by $A^R$) and the advanced (denoted by $A^A$)
averaged Green's function turn out to be,
$A^R=A^A=A_{++}$ for
$\kappa<(1+2v_{\text{i}}^{1/2})^{-1}$
or
$\kappa>(1-2v_{\text{i}}^{1/2})^{-1}$,
and for the interval
$(1+2v_{\text{i}}^{1/2})^{-1}<\kappa<(1-2v_{\text{i}}^{1/2})^{-1}$,
\begin{align}
 A^R
 &=
 \left\{
 \begin{array}{ll}
 A_{+-}\,\,& 
 (\mu>0)\\
 A_{++}\,\,& 
 (\mu<0)
 \end{array}
 \right. , 
 \label{eq:A_ret}\\
 A^A
 &=
 \left\{
 \begin{array}{ll}
 A_{++}\,\,& 
 (\mu>0)\\
 A_{+-}\,\,& 
 (\mu<0)
 \end{array}
 \right. .
 \label{eq:A_adv}
\end{align}
For a positive Fermi energy,
one of the eigenvalues of the averaged Green's functions multiplied by the Fermi energy
$\mu\tilde{G}_{k}$,
where the sign of the eigenvalue of $\tilde{\bm{\sigma}}\cdot \bm{k}$ is equal to
$\text{sgn}[\mu]$ (in this case +1),
is drawn in Fig. \ref{fig:GF} [(b), (c), (d)]
as a function of $\kappa=v_F\hbar|\bm{k}|/|\mu|$.
Each line in Fig. \ref{fig:GF} [(b), (c), (d)] corresponds to 
the intensity of disorder $v_{\text{i}}=2^{-n}\,(n=4,5,\cdots,10)$.
Fig. \ref{fig:GF} (a) shows the boundaries of the complex and the real
$A$ regions. 
At the boundaries, the Green's functions have singular behavior.
For a negative Fermi energy,
graphs are obtained by inverting the signs of the imaginary parts.
It is easily seen that in the low $v_{\text{i}}$ limit,
the graphs of the averaged Green's functions converge to those in the clean limit,
that is,
the real part converges to $(1-\kappa)^{-1}$,
and the imaginary part of the retarded [advanced] Green's function
to $-\pi\delta(\kappa-1)$ [$\pi\delta(\kappa-1)$],
since
\begin{align}
 \mu G_{0k}(\mu\pm i\delta)
 &=\frac{1}{1-\kappa\pm i\delta} \notag\\
 &=\frac{\text{P}}{1-\kappa}\mp i\pi\delta(\kappa-1).
\end{align}

Next, we consider the case $\mu=0$.
The algebraic equation of $A$ can be reduced from 
(\ref{eq:scEquation_A})
by taking $\mu=0$.
The equation is given by
\begin{align}
 A
 =
 -v_{\text{i}}(1-A)^2, 
 \label{eq:scEquation_A2}
\end{align}
and the solutions of (\ref{eq:scEquation_A2}) 
have two branches as
\begin{align}
 A_{\pm}=\frac{1}{2v_{\text{i}}}
 \left(
 2v_{\text{i}}-1\pm\sqrt{1-4v_{\text{i}}}
 \right).
 \label{eq:A_two_branch}
\end{align}
Since the solutions (\ref{eq:A_two_branch}) are real when
$v_{\text{i}}<0.25$,
we always consider real solutions of (\ref{eq:scEquation_A2}).
A branch of the solutions that converges to zero in the clean limit is 
$A_{+}$.
Note that the solutions (\ref{eq:A_two_branch}) are reduced from the solutions 
(\ref{eq:A_four_branch}) by taking
$(1\pm 2v_{\text{i}}^{1/2})\kappa\pm 1 \to (1\pm 2v_{\text{i}}^{1/2})\kappa$.

\subsection{Density of states}

The density of states of the Majorana surface modes at the energy $\epsilon$ is obtained
with the averaged Green's function  $\tilde{G}(\mu=\epsilon)$
by the formula
\begin{align}
 D(\epsilon)
 =
 -\frac{1}{\pi L^2}\sum_k
 \text{Im}\text{Tr}\,
 \tilde{G}_{k}(\epsilon+i\delta).
 \label{eq:DOS_def}
\end{align}
The sum over discrete $\bm{k}$ is replaced by the integration over continuous
$\bm{k}$ by taking the limit of $L\to \infty$: 
$\frac{1}{L^{2}}\sum_k\to\frac{1}{(2\pi)^{2}}\int d^2k$.
Substituting the averaged Green's function (\ref{eq:averaged_GF_A})
with the branch for $A^R$ given in (\ref{eq:A_ret})
into the above formula
and using the identity (\ref{eq:scEquation_A}),
the density of states is given by
\begin{align}
 D(\epsilon)
 &=
 -\frac{\epsilon}{\pi^2v_F^2}
 \int_0^{\infty}
 \frac{d\kappa}{v_{\text{i}}\kappa}
 \text{Im}
 \frac{A^R}{1+A^R} \notag\\
 &=
 \frac{|\epsilon|}{\pi^2v_F^2}
 \int_{(1+2v_{\text{i}}^{1/2})^{-1}}^{(1-2v_{\text{i}}^{1/2})^{-1}}
 \frac{d\kappa}{4v_{\text{i}}\kappa}
 (4v_{\text{i}}\kappa^2-(\kappa-1)^2)^{1/2}.
 \label{eq:DOS_2}
\end{align}
Here we have used the fact that, 
from the solutions of $A$ in (\ref{eq:A_four_branch}) with the signs shown in Fig. \ref{fig:A_region},
the imaginary part of the Green's function is nonzero only in the
interval
$(1+2v_{\text{i}}^{1/2})^{-1}<\kappa<(1-2v_{\text{i}}^{1/2})^{-1}$.
Since $A^R$ for $\mu>0$ and that for $\mu<0$ are related by
complex conjugation,
the imaginary part of the averaged Green's function for $\mu>0$ is equal to
that for $\mu<0$ multiplied by $(-1)$.
The sign of the imaginary part thus 
cancels the sign of $\epsilon$ in front of the integral in the first
line of (\ref{eq:DOS_2}).
Therefore the density of states is an even function of the energy
$\epsilon$,
and it depends only on the absolute value of the energy $|\epsilon|$.
The expression of the density of states is written by
\begin{align}
 D(\epsilon)
 &=
 \frac{|\epsilon|}{2\pi v_F^2}
 \frac{1-(1-4v_{\text{i}})^{1/2}}{2v_{\text{i}}(1-4v_{\text{i}})^{1/2}} 
 \notag\\
 &=
 \frac{|\epsilon|}{2\pi v_F^2}
 (1+3v_{\text{i}}+O(v_{\text{i}}^2)).
 \label{eq:DOS_imp}
\end{align}
 In the clean limit,
the density of states of the surface Majorana fermions
converges to $|\epsilon|/2\pi v_F^2$,
which is half of the density of states of $4\times 4$ Dirac fermions
systems, like graphene.
However, since two Majorana fermions are equivalent to a single complex
fermion,
the density of states of the complex fermions composed by the surface
Majorana fermions is 
quarter of that of the $4\times 4$ Dirac fermions.

\section{Thermal conductivity}

The electronic conductivity is obtained from the Green's
functions by the following formula,
\begin{align}
 \sigma(\epsilon)
 =
 \frac{1}{2}
 \text{Re}
 [I(\epsilon+i\delta,\epsilon-i\delta)-I(\epsilon+i\delta,\epsilon+i\delta)],
 \label{eq:conductivity_def2}
\end{align}
where
\begin{align}
 I(\epsilon,\epsilon')
 =
 \frac{e^2\hbar}{\pi L^2}
 \sum_k
 \text{Tr}
 \langle 
 v_xG_{k}(\epsilon)
 v_xG_{k}(\epsilon')
 \rangle,
 \label{eq:conductivity_I}
\end{align}
and the velocity operator is defined by
\begin{align}
 v_x=\frac{i}{\hbar}[H_0,x]=v_F\tilde{\sigma}^x.
\end{align}
The quantity $I(\epsilon,\epsilon')$ 
contains the disorder average of the product of two Green's functions
and $v_x$ between them:
$K(\epsilon,\epsilon')=\langle G(\epsilon)v_xG(\epsilon')\rangle$.
This is another quantity aside from the averaged Green's function to be calculated perturbatively.
Within the SCBA,
$K(\epsilon,\epsilon')$ is self-consistently determined with use of the 
averaged single Green's function as\cite{shon98}
\begin{align}
 K(\epsilon,\epsilon')
 =
 \tilde{G}(\epsilon)v_x\tilde{G}(\epsilon')
 +
 \tilde{G}(\epsilon)\langle
 UK(\epsilon,\epsilon')U\rangle\tilde{G}(\epsilon').
 \label{eq:scBorn_K1}
\end{align}
Then, the $k$ component of the above equation is as follows:
\begin{align}
 &K_k(\epsilon,\epsilon')
 =
 v_F
 \tilde{G}_k(\epsilon)\tilde{\sigma}^x\tilde{G}_k(\epsilon') \notag\\
 &\qquad+
 v_{\text{i}}
 \tilde{G}_k(\epsilon)
 (v_F\hbar\tilde{\bm{\sigma}}\cdot \bm{k})K_k(\epsilon,\epsilon')
 (v_F\hbar\tilde{\bm{\sigma}}\cdot \bm{k})
 \tilde{G}_k(\epsilon').
 \label{eq:scBorn_K2}
\end{align}
Iteratively substituting $K_k(\epsilon,\epsilon')$ in (\ref{eq:scBorn_K2}),
the formal solution of $K_k(\epsilon,\epsilon')$ is described by
 the sum of infinite series as
\begin{align}
 K_k(\epsilon,\epsilon')
 &=
 v_F
 \sum_{n=1}^{\infty}
 v_{\text{i}}^{n-1}
 \tilde{G}_k^{(n)}(\epsilon)\tilde{\sigma}^x\tilde{G}_k^{(n)}(\epsilon'),
 \label{eq:scBorn_K_infsum}
\end{align}
where
\begin{align}
 &\tilde{G}_k^{(1)}(\epsilon)=\tilde{G}_k(\epsilon), \\
 &\tilde{G}_k^{(n)}(\epsilon)=\tilde{G}_k(\epsilon)
 (v_F\hbar\tilde{\bm{\sigma}}\cdot \bm{k})\tilde{G}_k^{(n-1)}(\epsilon).
\end{align}
Introducing new variables $E$ and $\varphi$ by
$\epsilon/(1+A)=E\sinh \varphi$ and 
$v_F\hbar|\bm{k}|/(1+A)=E\cosh \varphi$,
and a matrix valued variable $s_k=\tilde{\bm{\sigma}}\cdot \bm{k}/|\bm{k}|$,
the averaged Green's function is written by
\begin{align}
 \tilde{G}_k(\epsilon)=-(\sinh\varphi+s_k\cosh\varphi)/E.
\end{align}
Then, $\tilde{G}_k^{(n)}(\epsilon)$ is recursively given by $E$ and
$\varphi$ as
\begin{align}
 \tilde{G}_k^{(n)}(\epsilon)
 =
 \frac{(v_F\hbar|\bm{k}|)^{n-1}}{(-E)^n}
 (\sinh[n\varphi]+s_k\cosh[n\varphi]).
\end{align}
The sum of the infinite series can be calculated with the identity
of the sum of power series 
$\sum_{n=1}^{\infty}x^n=x/(1-x)$,
as
\begin{align}
 K_k(\epsilon,\epsilon')
 =
 \frac{v_F}{4}
 &\left[
 \frac{\tilde{\sigma}^x+s_k\tilde{\sigma}^x+\tilde{\sigma}^xs_k+s_k\tilde{\sigma}^xs_k}
 {EE'e^{-\varphi-\varphi'}-v_{\text{i}}(v_F\hbar|\bm{k}|)^2} \right.\notag\\
 &+
 \frac{-\tilde{\sigma}^x-s_k\tilde{\sigma}^x+\tilde{\sigma}^xs_k+s_k\tilde{\sigma}^xs_k}
 {EE'e^{-\varphi+\varphi'}-v_{\text{i}}(v_F\hbar|\bm{k}|)^2} \notag\\
 &+
 \frac{-\tilde{\sigma}^x+s_k\tilde{\sigma}^x-\tilde{\sigma}^xs_k+s_k\tilde{\sigma}^xs_k}
 {EE'e^{\varphi-\varphi'}-v_{\text{i}}(v_F\hbar|\bm{k}|)^2} \notag\\
 &+
 \left.
 \frac{\tilde{\sigma}^x-s_k\tilde{\sigma}^x-\tilde{\sigma}^xs_k+s_k\tilde{\sigma}^xs_k}
 {EE'e^{\varphi+\varphi'}-v_{\text{i}}(v_F\hbar|\bm{k}|)^2} 
 \right],
\end{align}
where
$E$, $\varphi$ are for $\epsilon$,
and $E'$, $\varphi'$ are for $\epsilon'$.
After the subtraction in (\ref{eq:conductivity_def2}),
only the imaginary part of the inverse of the averaged Green's function
$Ee^{\pm\varphi}=\pm\epsilon/(1+A)+v_F\hbar|\bm{k}|/(1-A)$
contributes.
Thus we restrict our discussion to the case with complex $A$. 
The sum of infinite power series $\sum_{n=1}^{\infty}x^n$
 converges when $|x|<1$.
For $v_{\text{i}}<0.25$ and when $A$ is complex valued,
the sum appearing in (\ref{eq:scBorn_K_infsum}) does not converge 
since, for $\epsilon,\epsilon'>0$,
\begin{align}
 \frac{v_{\text{i}}(v_F\hbar|\bm{k}|)^2}{EE'e^{-\varphi-\varphi'}}=1
\end{align}
holds when $A$ for $\epsilon$ and $A$ for $\epsilon'$ are related by
complex conjugation, the case of which appears in 
$I(\epsilon+i\delta,\epsilon-i\delta)$.
The same result is true for $\epsilon,\epsilon'<0$
by replacing $e^{-\varphi-\varphi'}$ by $e^{\varphi+\varphi'}$.
Since no other terms that cancel the infinity appear,
we conclude that the conductivity away from $\epsilon=0$ is always infinity.
This result indicates that the conductivity is unaffected by the disorder
at $\epsilon\neq 0$.

Then, we proceed to the case $\epsilon=0$,
which is exactly the case of the surface of the topological superconductor. 
$\epsilon=0$ can be realized by taking the limit of $\varphi\to0$.
For $\epsilon=0\pm i\delta$, we obtain
\begin{align}
 Ee^{\varphi}\to\frac{v_F\hbar|\bm{k}|}{1-A}\pm i\delta, \quad
 Ee^{-\varphi}\to\frac{v_F\hbar|\bm{k}|}{1-A}\mp i\delta.
\end{align}
Here, we should note that
since the parameter $A$ is real for $\epsilon=0$,
we need an infinitesimal imaginary parameter $\pm i\delta$ to avoid the singularity.
The conductivity is then given by
\begin{align}
 &\sigma(0)
 =
 \frac{e^2v_F^2\hbar}{(2\pi)^2}
 \int_0^{\infty}kdk \notag\\
 &\qquad \times
 \left[
 \frac{2}{|v_F\hbar k/(1-A)+i\delta|^2-v_{\text{i}}(v_F\hbar k)^2}
 \right. \notag\\
 &\qquad \quad 
 -\frac{1}{(v_F\hbar k/(1-A)+i\delta)^2-v_{\text{i}}(v_F\hbar k)^2}
 \notag\\
 &\qquad \quad  \left.
 -\frac{1}{(v_F\hbar k/(1-A)-i\delta)^2-v_{\text{i}}(v_F\hbar k)^2}
 \right].
\end{align}
Finally we obtain the electronic conductivity at
$\epsilon=0$ as
\begin{align}
 \sigma(0)
 &=
 \frac{e^2}{\pi h}
 \frac{1}{4v_{\text{i}}^{1/2}}
 \left[
 \frac{1}{1/(1-A)-v_{\text{i}}^{1/2}}+
 \frac{1}{1/(1-A)+v_{\text{i}}^{1/2}}
 \right]
 \notag\\
 &\qquad \times 
 \log
 \left[
 \frac{1/(1-A)+v_{\text{i}}^{1/2}}{1/(1-A)-v_{\text{i}}^{1/2}}
 \right] 
 \notag\\
 &=
 \frac{e^2}{\pi h}
 \left[
 1
 +
 (10/3)v_{\text{i}}
 +
 O(v_{\text{i}}^2)
 \right].
\end{align}
In the clean limit,
the electronic conductivity of 
the surface Majorana modes converges to $e^2/\pi h$.

The electronic conductivity of Dirac fermions in the zero-energy limit 
is known to be a universal value of the order of $e^2/h$,
which is referred to as the minimal conductivity.
The minimal conductivity calculated from the Kubo formula
is sensitive to the order of taking limits of 
zero temperature, non perturbative (clean limit), 
and zero frequency (dc limit)\cite{ludwig94,ryu07,ziegler07}.
So far,
two coefficients of the minimal conductivity have been reported.
\begin{align}
 \sigma_1^{\text{min}}=\frac{1}{\pi}\frac{e^2}{h}, \\
 \sigma_2^{\text{min}}=\frac{\pi}{8} \frac{e^2}{h}.
\end{align}
When the dc limit is taken before the zero temperature limit and finally
the clean limit is taken, $\sigma_1^{\text{min}}$ is yielded.
Conversely, taking the clean limit before the zero-temperature limit and 
then taking the dc limit results in $\sigma_2^{\text{min}}$.
The minimal conductivity obtained in this paper 
is consistent with $\sigma_1^{\text{min}}$.

With the help of the Wiedemann-Franz law for the Majorana fermions,
the thermal conductivity of the surface of time-reversal-symmetric 
topological superconductors is given by
\begin{align}
 \kappa
 =
 \frac{1}{\pi}\frac{\pi^2k_B^2T}{6h}
 \left[
 1
 +
 (10/3)v_{\text{i}}
 +
 O(v_{\text{i}}^2)
 \right].
\end{align}
Note that the number of degrees of freedom that contribute to the
thermal conductivity is a quarter of $4\times 4$ Dirac fermions,
since the Hamiltonian is $2\times 2$ and the fermions are real (Majorana).

\section{Conclusion}

We have studied the disorder effects on the longitudinal thermal conductivity of the
Majorana surface modes of the three-dimensional time-reversal symmetric topological superconductor
within the SCBA.
Due to the two defining symmetries of the topological superconductor in
symmetry class DIII,
disorder appears in the Hamiltonian only as spatial deformations
 of the pair potential.
For the long-ranged limit of the Gaussian deformations around each scatterer,
the self-consistent Born equations are reduced to an algebraic equation
that can be exactly solved.

We have derived the density of states and the electric conductivity 
of the surface Majorana fermions by means of the Green's function technique.
The density of states is only modified by its coefficient,
while its dependence on the energy is unchanged.

The thermal conductivity is calculated from the electronic
conductivity via the Wiedemann-Franz law for Majorana fermions.
The electronic conductivity away from $\mu=0$
remains infinity, which means that it is unaffected by disorder
that is written by the gravitational field.
However, the electronic conductivity at $\mu=0$ (minimal conductivity),
which is realized in the surface of the topological superconductor
takes the finite value of the order of $e^2/h$.
In the clean limit, the minimal conductivity with a coefficient
$1/\pi$ appears.
The thermal conductivity in the clean limit is given by  
$(1/\pi)\cdot \pi^2k_B^2T/6h$.

\acknowledgements
The work of R. N. was supported by World Premier International Research Center
Initiative (WPI), MEXT, Japan. The work of K. N. was supported by MEXT Grant-in-Aid 
for Scientific Research (No. 24740211 and 25103703).

\end{document}